\documentclass[12pt]{article}
\pdfoutput=1

\usepackage{putex}

\usepackage{graphicx}
\usepackage{caption}
\usepackage{subcaption}
\usepackage{epstopdf}
\usepackage{enumerate}
\usepackage{cite}
\usepackage{tensor}
\usepackage{slashed}
\usepackage[aligntableaux=center]{ytableau}
\usepackage{bbm} 
\usepackage{amsthm}
\usepackage{array}
\usepackage{mathtools}
\usepackage{dsfont}
\usepackage{multirow}

\numberwithin{equation}{section}

\newcommand {\be} {\begin {equation}}
\newcommand {\ee} {\end {equation}}

\newcommand {\bes} {\begin {equation*}}
\newcommand {\ees} {\end {equation*}}

\newcommand{\es}[2] {\begin{equation} \label{#1} \begin{split} #2 \end{split} \end{equation}}

\newcommand{\Z}{\mathbb{Z}}

\newcommand{\beq}{\begin{equation}}
\newcommand{\eeq}{\end{equation}}

\def\<{\langle}
\def\>{\rangle}

\newcommand{\cO}{\ensuremath{\mathcal{O}}}

\begin{document}

\preprint{PUPT-2477}

\institution{Princeton}{Joseph Henry Laboratories, Princeton University, Princeton, NJ 08544, USA}

\title{Bootstrapping $O(N)$ Vector Models in $4<d<6$}

\authors{Shai~M.~Chester,\footnote{e-mail: {\tt schester@Princeton.EDU}} Silviu~S.~Pufu,\footnote{e-mail: {\tt spufu@Princeton.EDU}} and Ran~Yacoby\footnote{e-mail: {\tt ryacoby@Princeton.EDU}}}

\abstract{
We use the conformal bootstrap to study conformal field theories with $O(N)$ global symmetry in $d=5$ and $d=5.95$ spacetime dimensions that have a scalar operator $\phi_i$ transforming as an $O(N)$ vector.  The crossing symmetry of the four-point function of this $O(N)$ vector operator, along with unitarity assumptions, determine constraints on the scaling dimensions of conformal primary operators in the $\phi_i \times \phi_j$ OPE\@.  Imposing a lower bound on the second smallest scaling dimension of such an $O(N)$-singlet conformal primary, and varying the scaling dimension of the lowest one, we obtain an allowed region that exhibits a kink located very close to the interacting $O(N)$-symmetric CFT conjectured to exist recently by Fei, Giombi, and Klebanov.  Under reasonable assumptions on the dimension of the second lowest $O(N)$ singlet in the $\phi_i \times \phi_j$ OPE, we observe that this kink disappears in $d =5$ for small enough $N$, suggesting that in this case an interacting $O(N)$ CFT may cease to exist for $N$ below a certain critical value.
}

\date{December, 2014}

\maketitle

\tableofcontents

\setlength{\unitlength}{1mm}

\newpage
\section{Introduction and Summary}
\label{INTRO}	
The conformal bootstrap \cite{Polyakov:1974gs, Ferrara:1973yt, Mack:1975jr, Belavin:1984vu} has recently reemerged as a powerful tool for obtaining non-perturbative information about the operator spectrum and operator product expansion (OPE) coefficients of conformal field theories (CFTs).  Introduced originally in the context of two-dimensional CFTs \cite{Polyakov:1974gs, Ferrara:1973yt, Mack:1975jr, Belavin:1984vu}, this technique has been applied to higher-dimensional CFTs only recently starting with the work of \cite{Rattazzi:2008pe}. The bootstrap uses the crossing symmetry of correlation functions of CFTs to put an infinite set of constraints on the CFT data. While it is difficult to solve these constraints exactly in $d>2$, the recent reformulation of the bootstrap uses unitarity to rephrase the constraint problem as a convex optimization problem, which can be numerically solved to get bounds on the CFT data in any number $d$ of spacetime dimensions (see, for example, \cite{ Rattazzi:2008pe,Caracciolo:2009bx, Rychkov:2009ij, Rattazzi:2010gj, Poland:2010wg, Rattazzi:2010yc, Vichi:2011ux,Heemskerk:2010ty, Poland:2011ey,Rychkov:2011et, ElShowk:2012ht, ElShowk:2012hu, Liendo:2012hy,Kos:2013tga, El-Showk:2013nia,Gaiotto:2013nva, Beem:2013qxa, Beem:2013hha, Alday:2013opa, Alday:2013bha, Bashkirov:2013vya, Nakayama:2014lva, Nakayama:2014yia,Caracciolo:2014cxa, El-Showk:2014dwa, Berkooz:2014yda,Antipin:2014mga, Alday:2014qfa,Kos:2014bka,Chester:2014fya,Chester:2014mea, Nakayama:2014sba,Alday:2014tsa, Golden:2014oqa}). In several cases \cite{El-Showk:2014dwa,Golden:2014oqa,Kos:2013tga,Chester:2014mea}, these bounds have featured kinks that are believed to be located very close to known CFTs.  The conformal bootstrap has therefore allowed these known CFTs to be studied nonperturbatively.
	
A recent notable application of these techniques has been to CFTs with global $O(N)$ symmetry in $d=3$ \cite{Kos:2013tga}. This class of CFTs are of interest because they include the critical $O(N)$ vector model, which has several physical applications for $d=3$ and $N\leq3$. In $2<d<4$ the critical $O(N)$ vector model corresponds to the IR fixed point of the RG flow that starts from a free theory of $N$ scalars $\phi_i$, $i = 1, \ldots, N$, in the UV perturbed by the $(\phi_i \phi_i)^2$ operator. The critical $O(N)$ vector model in $2<d<4$ is strongly coupled in the IR, so perturbative analysis must employ either a large $N$ expansion \cite{Lang:1990ni,Lang:1991kp,Lang:1992zw,Lang:1992pp,Petkou:1995vu,Petkou:1994ad,Vasiliev:1982dc,Vasiliev:1981yc,Vasiliev:1981dg,Broadhurst:1996ur} or a Wilson-Fisher $\epsilon\ll1$ expansion for $d=4-\epsilon$ \cite{Wilson:1971dc,Wilson:1973jj}, neither of which extend easily to the physically interesting cases of small $N$ and $\epsilon=1$. The conformal bootstrap bounds on the scaling dimensions of CFTs with $O(N)$ symmetry feature a kink at values that seem to correspond to the perturbative prediction for the critical $O(N)$ vector model at large $N$ and are consistent with Monte Carlo simulation results at small $N$ \cite{Kos:2013tga}.  It is tantalizing that the conformal bootstrap may provide a new way of calculating the CFT data of the critical $O(N)$ model, and that this new way can perhaps have better accuracy than other current methods.  
	 
The conformal bootstrap for $O(N)$ theories successfully found the nontrivial critical theory in $d<4$; should we expect it to find another nontrivial interacting theory with $O(N)$ symmetry in $d>4$ spacetime dimensions? This question is particularly interesting due to the proposal that  $O(N)$ vector models in $d$ dimensions are dual to higher spin quantum gravity with Dirichlet boundary conditions in $d+1$ dimensions according to the AdS$_{d+1}$/CFT$_d$ correspondence \cite{Giombi:2012ms,Klebanov:2002ja}.  In $d>4$ spacetime dimensions, the quartic operator $(\phi_i \phi_i)^2$ is irrelevant, so there is no IR fixed point in this case that can be reached from the theory of $N$ free scalars perturbed by this quartic operator.  However, it has been proposed \cite{Parisi:1975im, Parisi:1977uz, Fei:2014yja,Fei:2014xta} that a nontrivial interacting $O(N)$ theory should also exist for $4<d<6$ as a UV fixed point from which one can flow to the theory of $N$ free scalars.  Such a UV fixed point can itself be thought of as the IR fixed point of the asymptotically free cubic theory of $N+1$ scalars with Lagrangian \cite{Giombi:2012ms,Klebanov:2002ja}
\es{lagrangian}{
  {\cal L}=\frac{1}{2}\left(\partial_\mu\phi_i\right)^2+\frac{1}{2}\left(\partial_\mu\sigma\right)^2+\frac{g_1}{2}\sigma\phi_i\phi_i+\frac{g_2}{2}\sigma^3.
}
Like in the $2<d<4$ version described above, when $4<d<6$, the nontrivial unitary IR fixed point of \eqref{lagrangian} is generically strongly coupled.    Perturbative analysis must employ either a large $N$ expansion or an $\epsilon\ll1$ expansion for $d=6-\epsilon$.\footnote{In the case of the $\epsilon$ expansion, it was found in \cite{Fei:2014yja,Fei:2014xta} that there are three non-trivial RG fixed points, two of which exist only for $N \gsim 1000$, while the latter exists for all values of $N$---See Figure~2 in~\cite{Fei:2014yja}.  It is not clear what the fate of these fixed points is as one increases $\epsilon$, but it is believed that at least one of them is still present when $\epsilon = 1$ and $N$ is large enough.}  Unlike the $2<d<4$ version, the $4<d<6$ interacting $O(N)$ vector model is predicted to have a critical value of $N$ below which the theory becomes non-unitary. The large $N$ expansion predicts that this $N_{\text{crit}}$ is small in $d=5$, roughly $N_{\text{crit}}<35$, so as in the $2<d<4$ case, the physically interesting regime is small $N$ and $\epsilon=1$, albeit for a different reason.  For such $N$ and $\epsilon$, the perturbative methods are inaccurate and the conformal bootstrap becomes one of the only available nonperturbative tools at our disposal.

The conformal bootstrap analysis of interacting $O(N)$ theories is more difficult in $d=5$ than in $d=3$ due to the following subtlety. In $d=3$, the critical theory was identified with a kink on the upper bound for the lowest dimension $\Delta_\sigma$ of an $O(N)$ singlet operator in terms of the dimension $\Delta_\phi$ of the $O(N)$ fundamental field $\phi_i$. This bound was derived from the crossing symmetry constraints on the 4-point function $\langle\phi_i\phi_j\phi_k\phi_l\rangle$ under no other assumptions besides unitarity for all other operators appearing in the $\phi_i \times \phi_j$ OPE\@.  Such single operator bootstrap studies with no further operator spectrum assumptions tend to produce monotonically increasing upper bounds that begin at value $(\Delta_\phi, \Delta_\sigma) = (\frac{d-2}{2}, d-2) $ corresponding to the free theory of $N$ scalars, at least when $\phi_i$ has small anomalous dimension.  (In the theory of $N$ free scalars, we have that $\sigma = \phi_i \phi_i$ is the lowest dimension $O(N)$ singlet.) The large $N$ expansion of the interacting $O(N)$ theory yields $\Delta_\phi = \frac{d-2}{2} + O(1/N)$ and $\Delta_\sigma = 2 + O(1/N)$ for all $d$.  Therefore, for $d=3$ the critical theory could exist as a kink on the border of the upper bound on $\Delta_\sigma$ as a function of $\Delta_\phi$, but for $d=5$ the interacting theory of \cite{Fei:2014yja,Fei:2014xta} would be hidden in the allowed region in the $(\Delta_\phi,\Delta_\sigma)$ plane below the free theory value $\Delta_\sigma = d-2$. 

One way of further carving out parts of the allowed region below $\Delta_\sigma = d-2$ is to examine more complicated 4-point functions. In the $O(1)=\mathbb{Z}_2$ Ising case, this region was probed by bootstrapping mixed correlators \cite{Kos:2014bka} and assuming that the only relevant scalar operators are $\phi$ and $\sigma$.  This assumption is physically motivated for the critical $O(N)$ vector models. These techniques yielded a small island of allowed region around the critical Ising point. Bootstrapping mixed correlators for general $O(N)$ theories is significantly more computationally intensive than for the $N=1$ case, however, because there are many more crossing relations for $N>1$. 

An alternative is to avoid the problem mentioned above altogether by using the conformal bootstrap to look at OPE coefficients instead of at scaling dimensions of operators.  Ref.~\cite{Nakayama:2014yia} determined bounds on the stress tensor and $O(N)$ current central charges,\footnote{These ``central charges'' are defined as the numerical coefficients that appear in the two-point functions of the canonically normalized stress tensor and $O(N)$ current, respectively.  If one normalizes the stress tensor and $O(N)$ current to a fixed number in all CFTs, the central charges mentioned above can be read off from the coefficients with which the stress tensor and of the $O(N)$ current appear in the $\phi_i \times \phi_j$ OPE.} $c_T$ and $c_J$, for $d=5$ $O(N)$-symmetric CFTs.  They found that, for large $N$, the minimum of $c_J$ seemed to correspond to the value expected from the large $N$ interacting theory value.  It was not clear, however, whether minimizing $c_J$ would match the interacting theory at smaller values of $N$.  Notably, in \cite{Nakayama:2014yia} no critical value of $N$ was identified.

In this paper, we find that simply imposing a lower bound on the {\em second lowest} conformal primary operator in the $O(N)$ singlet sector (which we will henceforth refer to as $\sigma^2$), and then employing the crossing symmetry of the single operator four point function $\langle\phi_i\phi_j\phi_k\phi_l\rangle$, along with unitarity assumptions, is enough to probe the region in the $(\Delta_\phi, \Delta_\sigma)$ plane below the free theory value $\Delta_\sigma = d-2$.\footnote{A similar observation has been made in $d=3$ for theories with a fermionic operator $\psi$ \cite{Fermions, PolandTalk}.}  We emphasize that the lower bound on $\Delta_{\sigma^2}$ (where this lower bound is chosen to be strictly greater than $\Delta_\sigma$) must be chosen judiciously if one's goal is to study the interacting $O(N)$ theories proposed in~\cite{Fei:2014yja,Fei:2014xta}.  Indeed, in these theories, one has $\Delta_{\sigma^2} = 4 + O(1/N)$, where the leading $1/N$ correction is negative  \cite{Lang:1990ni,Lang:1991kp,Lang:1992zw,Lang:1992pp,Petkou:1995vu,Petkou:1994ad,Vasiliev:1982dc,Vasiliev:1981yc,Vasiliev:1981dg,Broadhurst:1996ur}.  At least at large $N$, one therefore expects the scaling dimension of $\sigma^2$ to be slightly less than four.  Similarly, if one were to work in $d = 6-\epsilon$ dimensions, with $\epsilon \ll 1$, a perturbative computation shows that $\Delta_{\sigma^2} = 4 + O(\epsilon)$, where again the leading correction in $\epsilon$ is negative \cite{Fei:2014xta,Fei:2014yja}.  At least at small $\epsilon$, we therefore expect $\Delta_{\sigma^2}$ to be slightly less than four.  In this paper, we therefore choose the lower bound on $\Delta_{\sigma^2}$  to be slightly smaller than four, as informed by the perturbative expansions we just mentioned.

The summary of our paper is as follows.  We start with a study in $d=6-\epsilon$ space-time dimensions, with $\epsilon=.05$.  The advantage of such a study is that, in principle, we can be guided either by the $\epsilon$ expansion or by the $1/N$ expansion, or by both.\footnote{\label{FootnoteUnitarity}A slight disadvantage is that it has not been established conclusively whether the interacting $O(N)$ theories of~\cite{Fei:2014yja,Fei:2014xta} are unitary in non-integer dimensions.  Indeed, in \cite{Hogervorst:2014rta} it was noted that free field theories in fractional dimensions are non-unitary, and it was conjectured that the Wilson-Fisher fixed points would share the same feature.  In $d=4-\epsilon$, it can be checked explicitly that high-dimension operators acquire complex anomalous dimensions \cite{Rychkov}.    The numerical conformal bootstrap may not be sensitive to such mild potential violations of unitarity, as evidenced by the results of \cite{El-Showk:2013nia} on the critical $O(N)$ model in $2< d<4$.}  As we review in more detail in the following section, the $\epsilon$ expansion provides evidence for the existence of three RG fixed points in the model \eqref{lagrangian}, out of which only one is expected to be continuously connected to the interacting $O(N)$ model in $d=5$.  In $d=6 - \epsilon$ this fixed point exists only for values of $N$ larger than $N_\text{crit} \approx 1000$. Using the bootstrap, we find a kink for $N \gsim 1000$ whose $(\Delta_\phi,\Delta_\sigma)$ values match the $\epsilon\ll1$ perturbative prediction for this fixed point. Interestingly, we find that this kink persists, however, for $N \lsim 1000$ as well, and its existence does not seem to be related to one of the other RG fixed points that is expected to exist for all $N$.

We continue with a study of $O(N)$-symmetric theories in $d=5$.  We start with $N=500$ and using the large $N$ expansion value for $\Delta_{\sigma^2}$ and our most accurate numerics we find a kink at $(\Delta_\phi,\Delta_\sigma)=(1.500409,2.027)$ that is very close to the large $N$ expansion values $(\Delta_\phi,\Delta_\sigma)=(1.500414,2.022)$ for the critical theory.  We then examine $N\leq40$ and find a kink that disappears around $15<N_\text{crit}<22$ for a reasonable assumption of $\Delta_{\sigma^2}$, which is consistent with the large $N$ expansion prediction of $N_{\text{crit}}<35$. We check that our choice of $\Delta_{\sigma^2}$ does not qualitatively affect our answers for $6\leq N\leq40$.

The outline of this paper is as follows. In Section~\ref{critical theory} we briefly list some relevant facts about the recently proposed $4<d<6$ $O(N)$-symmetric critical theory in $d=5$ as well as $d=6-\epsilon$ for $\epsilon=.05$. In Section~\ref{numerics} we present our bootstrap bounds for $d=5.95$ and $d=5$. Lastly, Section~\ref{discussion} contains an interpretation of our results with further discussion.

{\bf Note added:}  As this work was in its final stages, the preprint \cite{Bae:2014hia}, which explores the same topic as we do, appeared on {\tt arXiv.org}.


\section{Review of Nontrivial $O(N)$-symmetric  Theories for $4<d<6$ }
\label{critical theory}	
The large $N$ analysis of the interacting $O(N)$ vector model has been carried out for arbitrary dimension $d$ in \cite{Lang:1990ni,Lang:1991kp,Lang:1992zw,Lang:1992pp,Petkou:1995vu,Petkou:1994ad,Vasiliev:1982dc,Vasiliev:1981yc,Vasiliev:1981dg,Broadhurst:1996ur}. For the two cases studied in this work, $d=5$ and  $d=5.95$, we record that the dimensions for the $O(N)$ fundamental $\phi_i$ and the two lowest singlet operators $\sigma$ and $\sigma^2$ are
\begin{subequations}
\label{5}
\begin{align}
d=5: \qquad \Delta_{\phi}&=1.5+\frac{0.216152}{N}-\frac{4.342}{N^2}-\frac{121.673}{N^3}+O(1/N^4) \label{philargeN5}  \,, \\
\label{sigmalargeN5}
\Delta_\sigma&=2+\frac{10.3753}{N}+\frac{206.542}{N^2}+O(1/N^3) \,, \\
\Delta_{\sigma^2}&=4-\frac{13.8337}{N}-\frac{1819.66}{N^2}+O(1/N^3) \,,
\label{sigma2largeN5} 
\end{align}
\end{subequations}
\begin{subequations} \label{5.95}
\begin{align}
\qquad\quad d=5.95:\qquad \Delta_{\phi}&=1.975+\frac{.0476989}{N}-\frac{1.86705}{N^2}-\frac{58.7662}{N^3}+O(1/N^4) \,, \label{philargeN}\\
\label{sigmalargeN}
\Delta_\sigma&=2+\frac{1.91309}{N}+\frac{311.496}{N^2}+O(1/N^3) \,, \\
\Delta_{\sigma^2}&=4-\frac{4.67497}{N}-\frac{2287.35}{N^2}+O(1/N^3) \,.
\label{sigma2largeN}
\end{align}
\end{subequations}
For $d=5$, one can estimate that $\Delta_{\phi}$ no longer obeys the unitarity bound $\Delta_{\phi}\geq3/2$ if $N$ is smaller than $N_{\text{crit}} \approx 35$, although this estimate is of course very crude \cite{Fei:2014xta,Fei:2014yja}.

In $d=6-\epsilon$, one can also use the $\epsilon\ll1$ expansion computed in \cite{Fei:2014xta,Fei:2014yja} for the Lagrangian \eqref{lagrangian}. In that study, the $\beta$-function $\beta(g_1,g_2)$ and the anomalous dimensions $\gamma_\sigma(g_1,g_2)$ and $\gamma_\phi(g_1,g_2)$ were computed to three loop order $O(\epsilon^3)$, while the anomalous dimension $\gamma_{\sigma^2}(g_1,g_2)$ was computed to one loop order $O(\epsilon)$. The fixed point couplings $g_1^*, g_2^*$ are then found by solving for $\beta(g_1^*,g_2^*)=0$, and can then be plugged into the anomalous dimensions to find the scaling dimensions at the fixed points. In general there are three distinct solutions $(g_1^*,g_2^*)$: a solution that corresponds to the interacting $O(N)$ model with two relevant scalar operators, and two solutions that correspond to theories with three relevant scalar operators. In the Appendix we list the $\epsilon$ expansion for each solution for the various values of $N$ used in this paper for $d=5.95$. The interacting $O(N)$ fixed point coupling solution becomes complex at
\es{Ncrit}{
N_{\text{crit}}=1038.26605-609.83980\epsilon-364.17333\epsilon^2+O(\epsilon^3) \,.
}
Note that while one of the solutions with three relevant scalar operators also becomes complex at this point, the other one remains real until
\es{Ncrit2}{
N'_{\text{crit}}=1.02145+0.03253\epsilon-0.00163\epsilon^2+O(\epsilon^3) \,,
}
i.e. effectively for all $N>0$.


\section{Conformal Bootstrap Numerics}
\label{numerics}
Let us briefly review the formulation of the numerical conformal bootstrap for CFTs with $O(N)$ global symmetry.  For further details, see \cite{Kos:2013tga}. Invariance of the four point function of $O(N)$ fundamental fields
\es{4pt}{
\langle\phi_i(x_1)\phi_j(x_2)\phi_k(x_3)\phi_l(x_4)\rangle
}
under the exchange $(x_1,i)\leftrightarrow(x_3,k)$ implies the crossing equation
\es{crossing}{
  \sum_{\cO \in\, \phi_i \times \phi_j \text{ OPE}} \lambda^2_{\cO}\, \vec{d}_{R, J}(\Delta_{\cO},\Delta_\phi)=0 \,, 
}
where $\cO $ runs over all conformal primaries in the OPE $\phi_i\times\phi_j$.  Here, $\lambda^2_{\cO}$ are the squares of the OPE coefficients that must be positive by unitarity, and $\vec{d}_{R, J}(\Delta_{\cO},\Delta_\phi)$ are explicit functions of the conformally-invariant cross-ratios $u = \frac{x_{12}^2 x_{34}^2}{x_{13}^2 x_{24}^2}$ and $v = \frac{x_{14}^2 x_{34}^2}{x_{23}^2 x_{24}^2}$ whose form depends only on the dimension of both the $O(N)$ fundamental $\Delta_{\phi}$ and on the dimension $\Delta_{\cO}$, Lorentz spin $J$, and $O(N)$ irrep $R$ of the operator $\cO$.  ($R \in \{s, t, a\}$, where $s$, $t$, $a$ represent $O(N)$ singlets, rank-two symmetric traceless tensors, and rank-two anti-symmetric tensors, respectively.) As in \cite{Kos:2013tga}, we normalize the OPE coefficient of the identity operator $\lambda_\text{Id}$=1.

To find bounds on the scaling dimensions of operators appearing in the $\phi_i \times \phi_j$ OPE, one can consider linear functionals $\alpha$ satisfying the following conditions:
\es{bootstrapOld}{
&\alpha(\vec{d}_{s, 0}(0, \Delta_\phi))=1 \,, \\
&\alpha(\vec{d}_{R, J}(\Delta ,\Delta_\phi))\geq0, \quad \text{for all $\Delta \geq\Delta_{R, J}^*$} \\
}
where $\Delta^*_{R, J}$ are the assumed lower bounds for spin-$J$ conformal primaries (other than the identity) that appear in the $\phi_i \times \phi_j$ OPE and transform in the $O(N)$ irrep $R$. The existence of any such $\alpha$ would contradict \eqref{crossing}, and thereby would allow us to find a combined upper bound on the lowest-dimension $\Delta^*_{R, J}$ of the spin-$J$ conformal primary in irrep $R$.  If we set $\Delta^*_{s, 0} = \Delta_\sigma$ and all other $\Delta^*_{R, J}$ equal to the corresponding unitarity value, we can then find disallowed points in the $(\Delta_\phi,\Delta_\sigma)$ plane.  As mentioned in the introduction, this procedure gives a monotonically increasing upper bound for $\Delta_\sigma$ vs.~$\Delta_\phi$, and the interacting theories discussed in $4<d<6$ dimensions discussed in the previous section sit well within the allowed region.

To overcome this difficulty, we modify the procedure mentioned above and look for functionals $\alpha$ satisfying
\es{bootstrapNew}{
&\alpha(\vec{d}_{s, 0}(0, \Delta_\phi))=1 \,, \\
&\alpha(\vec{d}_{s, 0}(\Delta_\sigma ,\Delta_\phi))\geq0 \,, \\
&\alpha(\vec{d}_{s, 0}(\Delta ,\Delta_\phi))\geq0, \quad \text{for all $\Delta \geq \Delta^*_{s, 0} > \Delta_\sigma$} \,, \\
&\alpha(\vec{d}_{R, J}(\Delta ,\Delta_\phi))\geq0, \quad \text{for all $(R, j) \neq (s, 0)$ and $\Delta \geq \Delta^*_{R, J}$} \,. \\
}
The existence of such a functional $\alpha$ disproves the existence of an $O(N)$-symmetric SCFT for which the conformal primaries appearing in the $\phi_i \times \phi_j$ OPE satisfy the following conditions:  
 \begin{itemize}
  \item The lowest spin-$0$ $O(N)$ singlet has dimension $\Delta_\sigma$.
  \item All other spin-$0$ $O(N)$ singlets have dimensions larger than $\Delta^*_{s, 0}$.
  \item All spin-$J$ conformal primaries other than spin-$0$ Lorentz singlets have dimension larger than $\Delta^*_{R, J}$.  
 \end{itemize}
From now on we set $\Delta^*_{R, J}$ equal to the unitarity bound for all $(R, J) \neq (s, 0)$, and we interpret $\Delta^*_{s, 0}$ as a lower bound on the second lowest dimension $\Delta_{\sigma^2}$ of a spin-$0$ $O(N)$ singlet conformal primary.  We denote this second lowest dimension by $\Delta_{\sigma^2}$ because in the interacting $O(N)$-symmetric theory described by the Lagrangian \eqref{lagrangian} the corresponding operator is $\sigma^2$.

The numerical implementation of the above problem requires two truncations: one in the number of derivatives used to construct $\alpha$ and one in the range of spins $J$ that we consider, whose contributions to the conformal blocks are exponentially suppressed for large spin  $J$.  We denote the maximum derivative order by $\Lambda$ (as in \cite{Chester:2014fya}) and the maximum spin by $J_\text{max}$.  The truncated constraint problem can then be rephrased as a semidefinite programing problem using the method developed in \cite{Rattazzi:2008pe}. This problem can be solved efficiently by freely available software such as {\tt sdpa\_gmp} \cite{sdpa}. The limiting factor in this implementation of the numerics is the parameter $\Lambda$. In this study we were able to compute numerically stable results for $\Lambda\leq21$ and spins up to $J_\text{max} = 30$.  In each of the following cases we specify what values of $\Lambda$ and $J_\text{max}$ were used.

\subsection{Bounds for $d=5.95$}
\label{5.95d}

\begin{figure}[t!]
\begin{center}
   \includegraphics[width=0.7\textwidth]{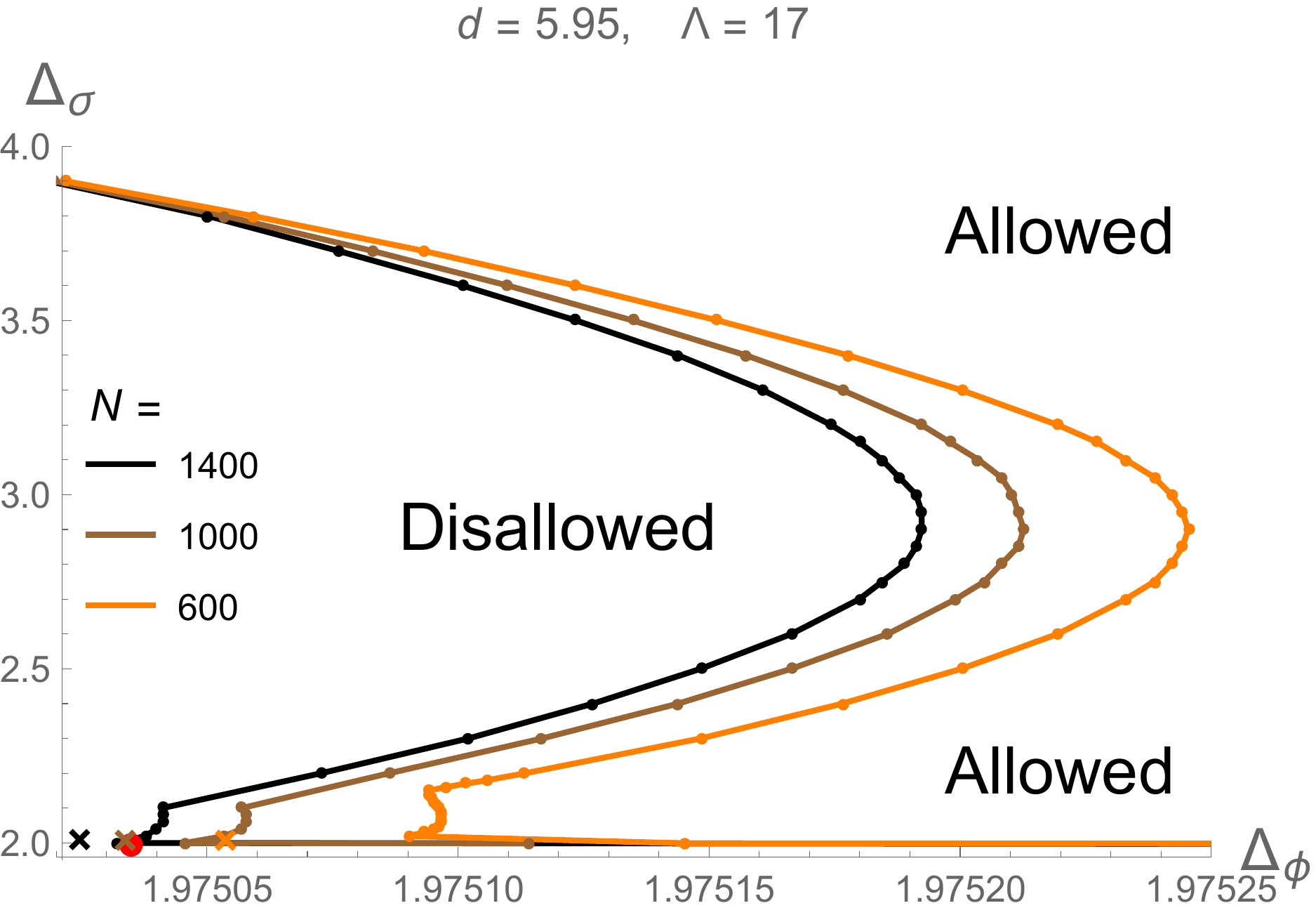}
\caption{Bounds on $\Delta_\sigma$ in terms of $\Delta_\phi$ in $d=5.95$ for $N=600,1000,1400$ under the assumption that $\sigma$ is the only scalar operator with dimension less than $\Delta_{\sigma^2}\geq3.986, 3.993, 3.996$ respectively. These bounds were computed with $J_\text{max}=20$ and $\Lambda=17$. The red dot denotes the large $N$ expansion $(\Delta_\phi,\Delta_\sigma)$ values for the critical $O(N)$ vector model for $N=1400$. The crosses denote the $\epsilon$ expansion $(\Delta_\phi,\Delta_\sigma)$ values for the CFT with three relevant operators that exists for all $N>0$.}
\label{fig:N595global}
\end{center}
\end{figure}

In Figure~\ref{fig:N595global} we show bounds on $\Delta_\sigma$ in terms of $\Delta_\phi$ in $d = 5.95$ for $N$ near $(N=1000)$ the large $N$ expansion $N_\text{crit}\approx 1000$, as well as above $(N=1400)$ and below $(N=600)$. In obtaining these plots we used $\Delta_{\sigma^2}=3.993, 3.996, 3.986$, respectively, as obtained from the large $N$ expansion \eqref{5.95}.  Interestingly, these plots show an allowed region that starts at the free theory point $(\Delta_\phi, \Delta_\sigma) = (1.975, 3.95)$ and that exhibits a kink close to $\Delta_\sigma \approx 2$, as expected for the RG fixed points mentioned in the previous section.  

Recall that in the $\epsilon$ expansion around $d=6$, there are three RG fixed points, two of which exist only for $N$ larger than $N_\text{crit}$ and one that exists for all values of $N$.  When $N=1400$, the values of $(\Delta_\phi, \Delta_\sigma)$ corresponding to the two fixed points that exist only at large $N$ are numerically very close and are marked with a red dot in Figure~\ref{fig:N595global}.  They are also very close to the kink mentioned in the previous paragraph, which suggests that this kink corresponds to one of these two interacting theories in $d=5.95$.  The values of $(\Delta_\phi, \Delta_\sigma)$ for the fixed point that exists at all $N$ are marked with crosses of different colors in Figure~\ref{fig:N595global}.  They are all in the disallowed region, perhaps due to the fact that the values of $\Delta_{\sigma^2}$ for these theories are significantly below the lower bounds on $\Delta_{\sigma^2}$ we used in making these plots.

It is worth noting that the kink close to $\Delta_\sigma \approx 2$ persists even below $N_\text{crit}$.  If one were to continue the RG fixed points that exist only above $N_\text{crit}$ to smaller values of $N$, one would obtain fixed points whose critical couplings acquire imaginary parts.  These fixed points therefore continue as non-unitary theories for $N < N_\text{crit}$.\footnote{As mentioned in Footnote~\ref{FootnoteUnitarity}, it is not clear whether the interacting $O(N)$ theories of \cite{Fei:2014yja,Fei:2014xta} are unitary in non-integer dimension even at large $N$.}  We can speculate that the kinks in Figure~\ref{fig:N595global} for $N\leq 1000$ are linked to these non-unitary theories, where the violations of unitarity are too small to be detected by our numerics.  It would be very interesting to find a way of reproducing $N_\text{crit} \approx 1000$ from a bootstrap computation, and we leave this question open for future work.

\begin{figure}[t!]
\begin{center}
   \includegraphics[width=0.7\textwidth]{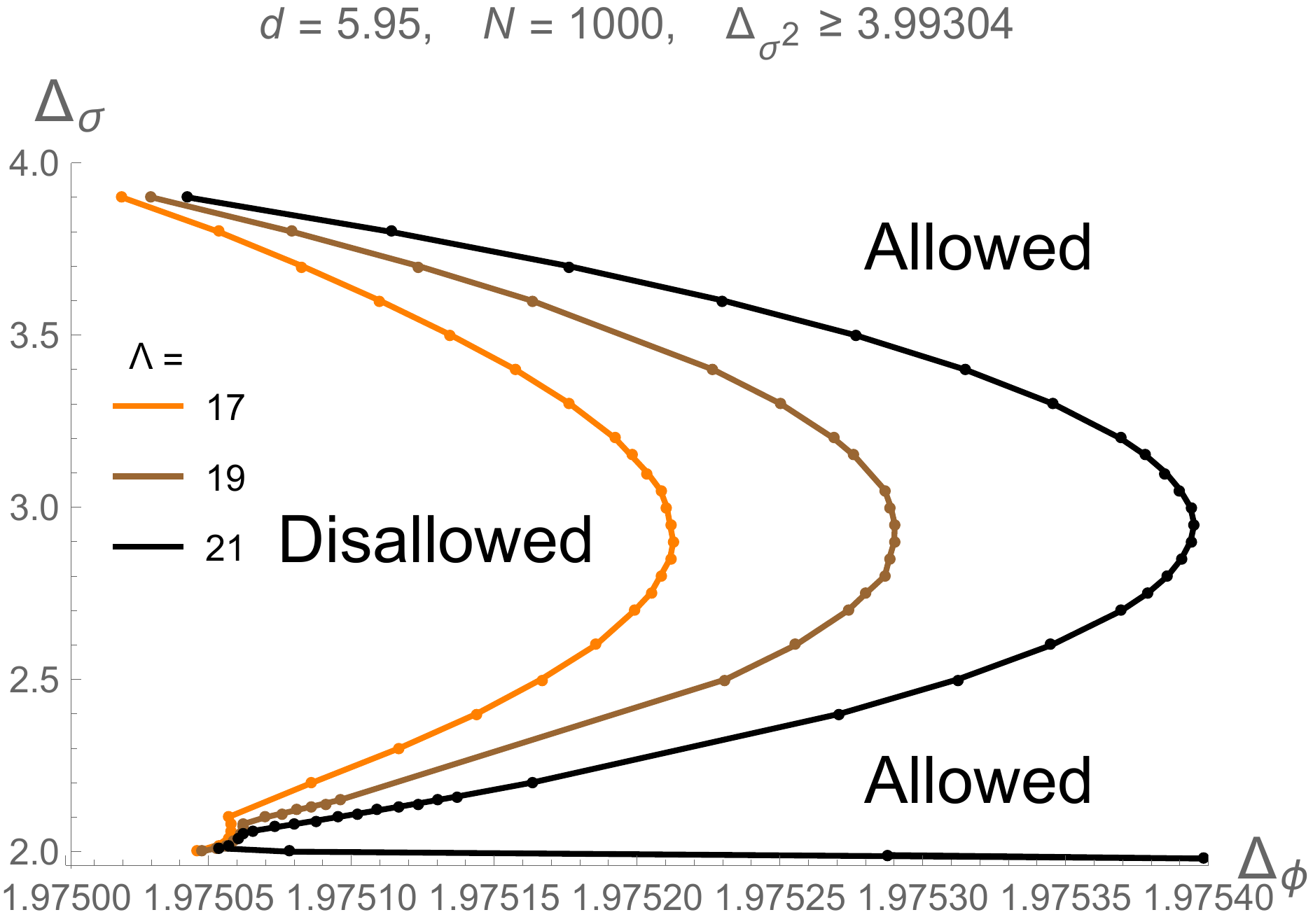}
\caption{Bounds on $\Delta_\sigma$ in terms of $\Delta_\phi$ in $d=5.95$ for $N=1000$ under the assumption that $\sigma$ is the only scalar operator with dimension less than $\Delta_{\sigma^2}\geq3.993$. The black line was computed with $J_\text{max}=30$ and $\Lambda=21$, the brown line was computed with $J_\text{max}=25$ and $\Lambda=19$, and the orange line was computed with $J_\text{max}=20$ and $\Lambda=17$. Note that the lower kink corresponding to the interacting $O(N)$ CFT is well converged, but the second higher kink diminishes significantly as $\Lambda$ is increased.}
\label{fig:595Compare}
\end{center}
\end{figure}

It is also worth noting that in Figrure~\ref{fig:N595global} there is a less pronounced second kink close to the kink at $\Delta_\sigma \approx 2$ we just discussed.  However, this second kink becomes less and less pronounced as we increase the parameter $\Lambda$ in our numerics---see Figure~\ref{fig:595Compare} for a comparison between $\Lambda = 17$, $\Lambda = 19$, and $\Lambda = 21$ when $N=1000$.  We notice that the disallowed region becomes significantly larger as we increase $\Lambda$, but the main kink around $\Delta_\sigma \approx 2$ discussed above is rather well-converged.  The same phenomenon is observed also for $N=600$ and $N=1400$.

\subsection{Bounds for $d=5$}
\label{5d}

\begin{figure}[t!]
\begin{center}
   \includegraphics[width=0.7\textwidth]{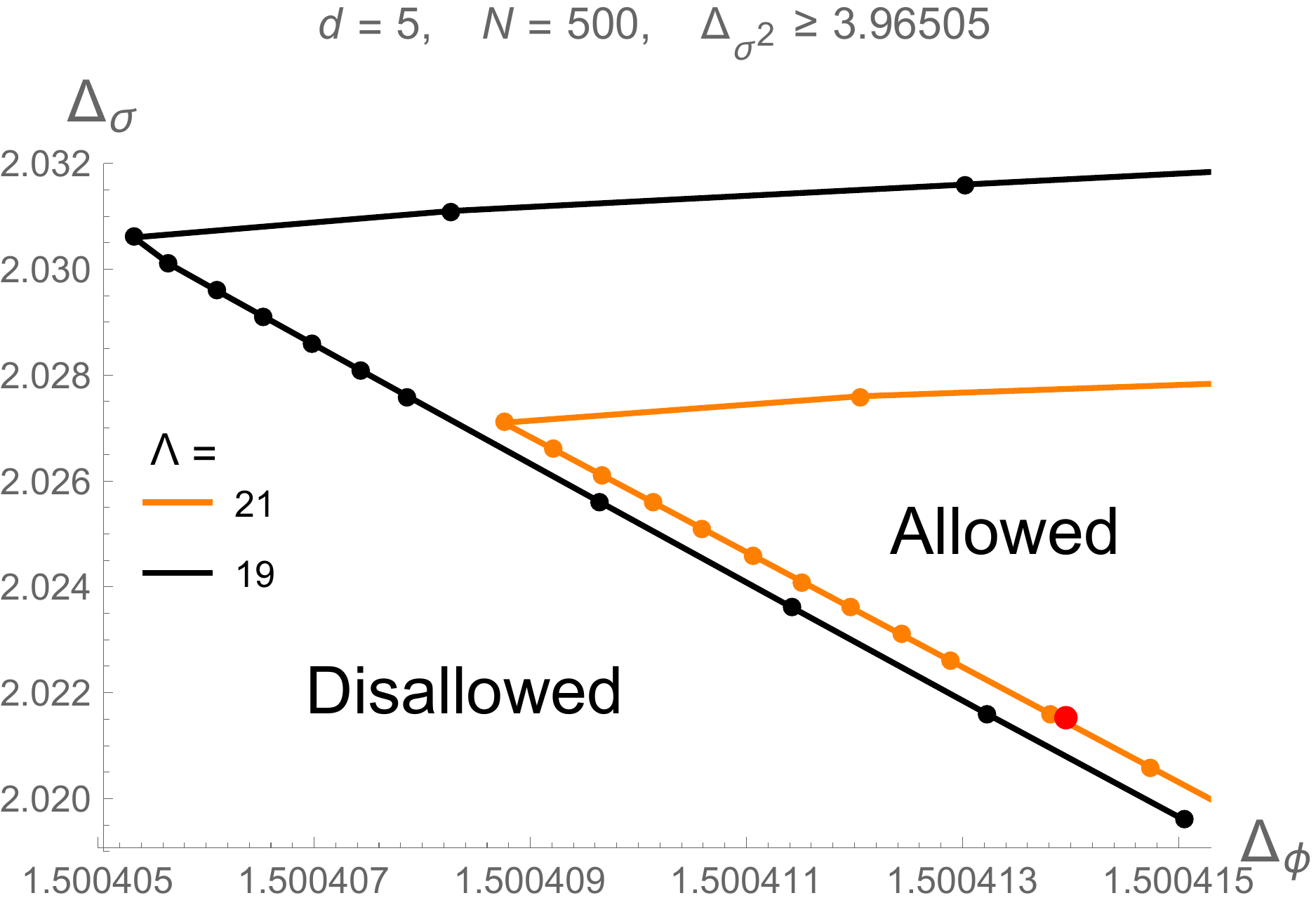}
\caption{Bounds on $\Delta_\sigma$ in terms of $\Delta_\phi$ in $d=5$ for $N=500$ under the assumption that $\sigma$ is the only scalar operator with dimension less than $\Delta_{\sigma^2}\geq3.965$. The black line was computed with $J_\text{max}=25$ and $\Lambda=19$, while the orange line was computed with $J_\text{max}=30$ and $\Lambda=21$. The red dot denotes the large $N$ expansion $(\Delta_\phi,\Delta_\sigma)=(1.500414,2.022)$ for the critical $O(N)$ vector model. Note the extremely zoomed in scale of this plot.}
\label{fig:N500}
\end{center}
\end{figure}

Let us now show numerical bootstrap bounds in $d=5$.  In Figure~\ref{fig:N500} we show bounds on $\Delta_\sigma$ in terms of $\Delta_\phi$ for $N=500$, with $J_\text{max}=25, 30$ and $\Lambda=19, 21$ respectively.   The value $N=500$ is large enough that we can use the $1/N$ expansion in \eqref{5} reliably.  In particular, we use $\Delta_{\sigma^2} = 3.965$ as an accurate lower bound for our computation.  As in the $d=5.95$ plots described above, the global shape of the allowed region in the $(\Delta_\phi, \Delta_\sigma)$ plane shows a kink below the free theory value $\Delta_{\sigma} = 3$.  For $\Lambda = 21$, this kink is located at $(\Delta_\phi,\Delta_\sigma)=(1.500409,2.027)$, which matches the large $N$ values computed from \eqref{5} rather well. (To the accuracy in \eqref{5}, we have $(\Delta_\phi, \Delta_\sigma) \approx (1.500414, 2.022)$.) Note that as we increase $\Lambda$, the kink seems to roughly move along the lower border of the allowed region, which does not change significantly.  It is encouraging that the analytical approximation marked by a red dot in Figure~\ref{fig:N500} also lies very close to the lower border of the allowed region, because it is likely that as we increase $\Lambda$ the kink would move closer to the red dot.

\begin{figure}[t!]
\begin{center}
   \includegraphics[width=0.7\textwidth]{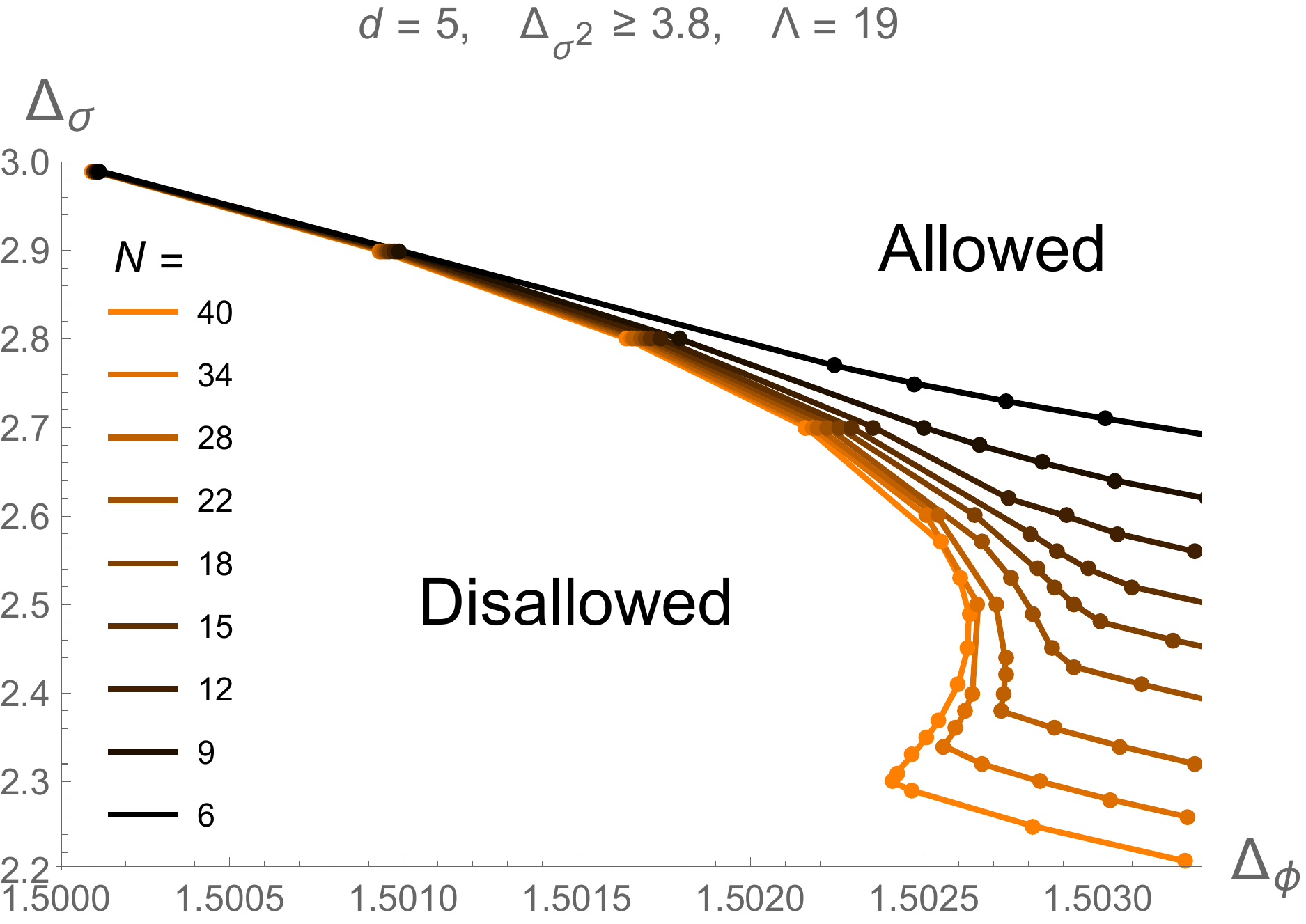}
\caption{Bounds on $\Delta_\sigma$ in terms of $\Delta_\phi$ in $d=5$ for a range of $N$ under the assumption that $\sigma$ is the only scalar operator with dimension less than $\Delta_{\sigma^2}\geq3.8$. These bounds are computed with $J_\text{max}=25$ and $\Lambda=19$.}
\label{fig:5dLowN}
\end{center}
\end{figure}

In Figure~\ref{fig:5dLowN} we show bounds on $\Delta_\sigma$ in terms of $\Delta_\phi$ for a range of $N$ near $N_{\text{crit}}\approx35$ as approximated by the large $N$ expansion. These plots were made under the assumption that $\sigma$ is the only scalar operator with dimension less than $\Delta_{\sigma^2}\geq3.8$.  For this plot we used $J_\text{max}=25$ and $\Lambda=19$. The most notable features of this plot are the sharp kinks clearly noticeable for the larger values of $N$, which seem to disappear between $N=15$ and $N=22$. 

\begin{figure}[t!]
\begin{center}
   \includegraphics[width=0.49\textwidth]{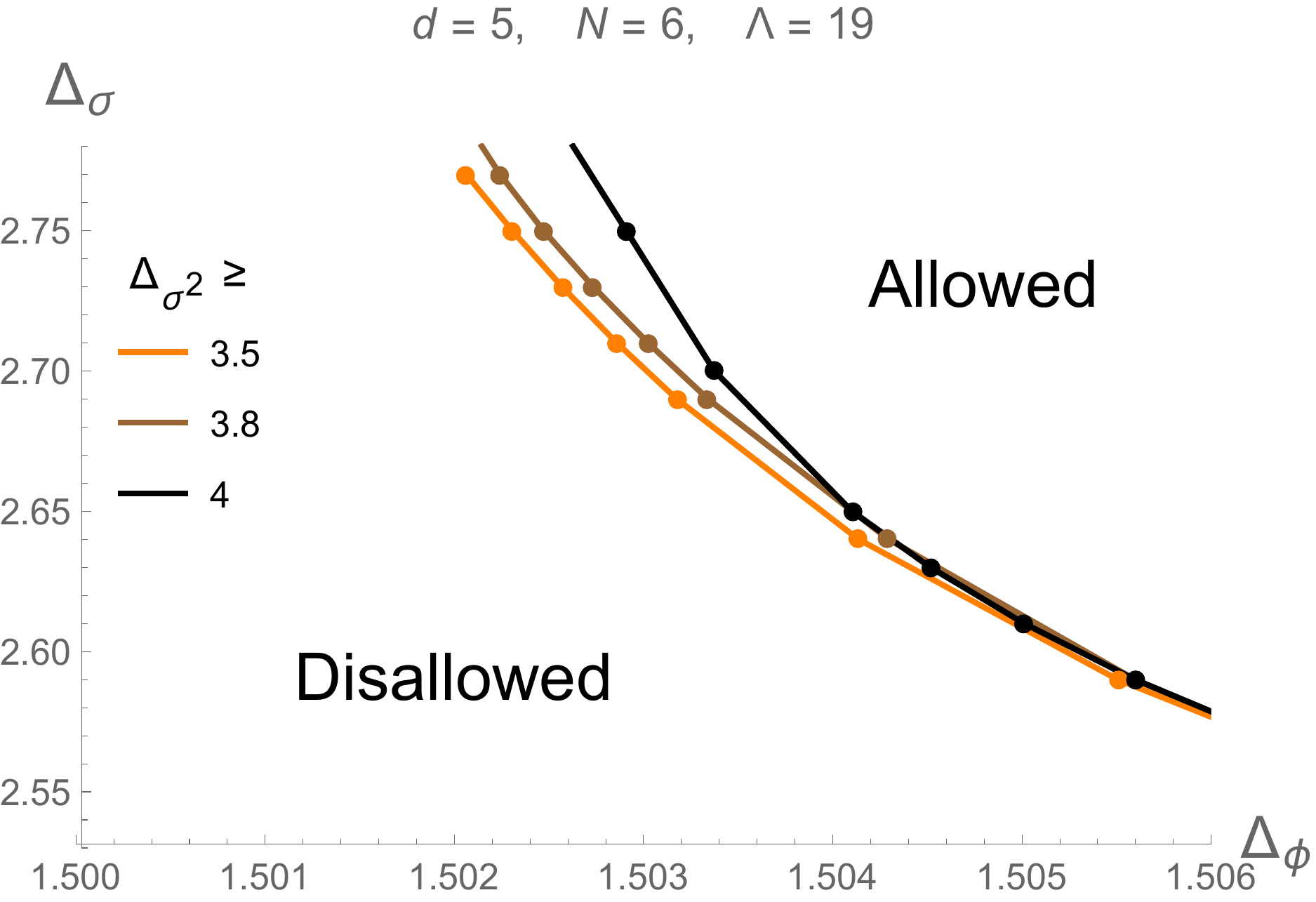}
   \includegraphics[width=0.5\textwidth]{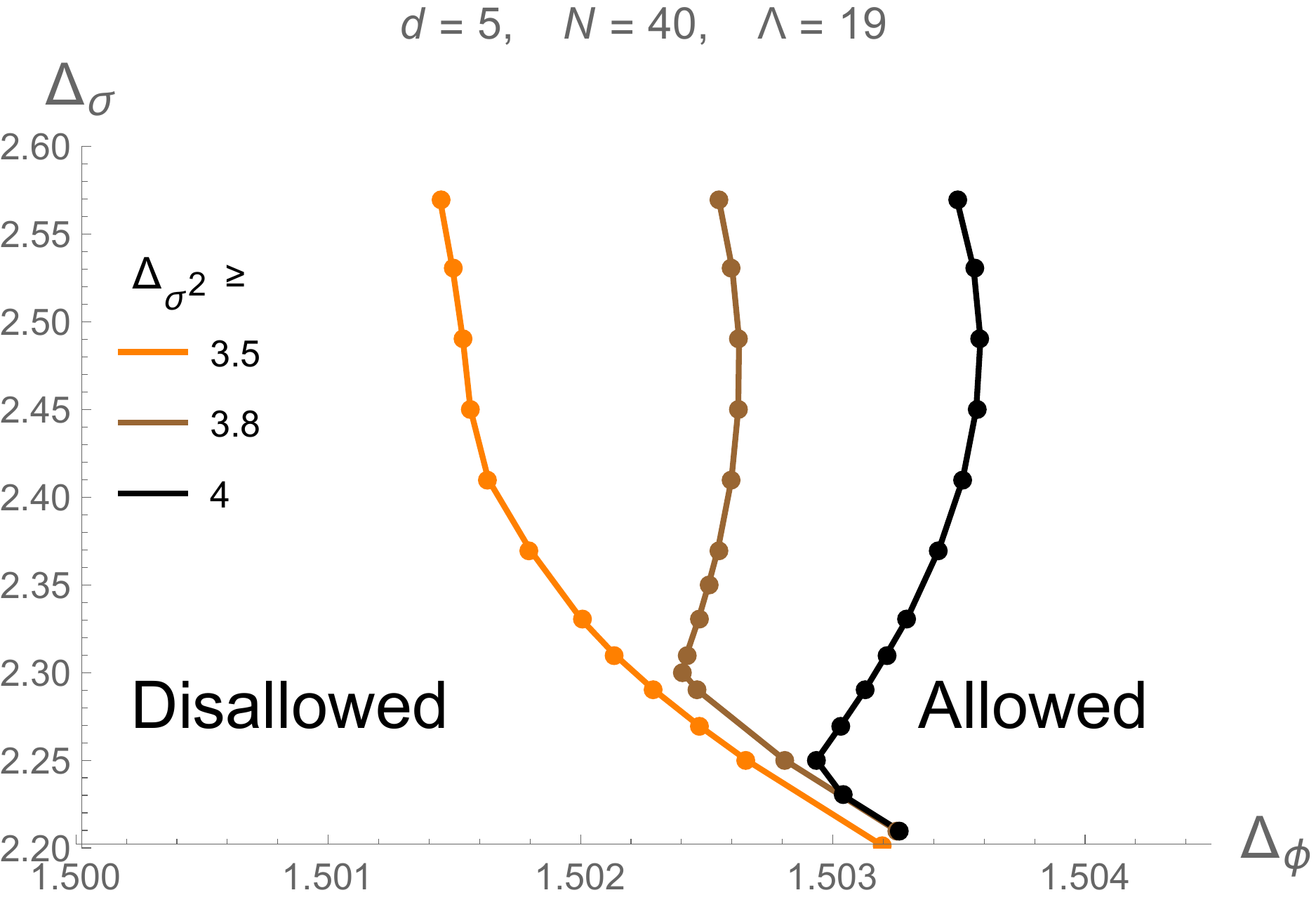}
\caption{Bounds on $\Delta_\sigma$ in terms of $\Delta_\phi$ in $d=5$ for $N=6$ (left) and $N=40$ (right) under the assumption that $\sigma$ is the only scalar operator with dimension less than $\Delta_{\sigma^2}$. The solid lines were computed with $J_\text{max}=25$ and $\Lambda=19$ for a variety of assumed lower bounds for $\Delta_{\sigma^2}$.}
\label{fig:5dgaps}
\end{center}
\end{figure}

At the low values of $N$ considered in Figure~\ref{fig:5dLowN} the large $N$ expansion for $\Delta_{\sigma^2}$ \eqref{sigma2largeN5} no longer provides an accurate estimate for the lower bound on $\Delta_{\sigma^2}$. We chose $\Delta_{\sigma^2}\geq3.8$ as a reasonable estimate, considering that at $N>200$ the large $N$ expansion is reasonably accurate, gives $\Delta_{\sigma^2}>3.9$, and tends to be monotonically increasing with $N$ fairly slowly. In Figure~\ref{fig:5dgaps} we show that considering large or smaller bounds for $\Delta_{\sigma^2}$ does not qualitatively change the features in our plots, in particular the appearance or lack of appearance of a kink, for the highest $(N=40)$ and lowest $(N=6)$ values of $N$ that we consider.   Note that a kink does not appear if we assume a less stringent bound on $\Delta_{\sigma^2}$, so we can say reliably that our plots feature no kinks for small values of $N$ and physical values of $\Delta_{\sigma^2}$.  (Recall that we expect $\Delta_{\sigma^2}$ to be less than $4$ for the interacting $O(N)$ models proposed in \cite{Fei:2014yja,Fei:2014xta}.) These plots were computed for $J_\text{max}=25$ and $\Lambda=19$.

\section{Discussion}
\label{discussion}

We end with a discussion of our results.  Our main observation in this work is that one can study interacting $O(N)$-symmetric CFTs in more than four spacetime dimensions using the conformal bootstrap, and that the conformal bootstrap provides evidence for the RG fixed points found in~\cite{Fei:2014yja,Fei:2014xta}.  Moreover, we provide evidence that the conformal bootstrap can lead to a precision study of these theories, as they seem to be located near sharp features (kinks) in the space of allowed scaling dimensions for certain operators.  More explicitly, our strategy is to consider the $O(N)$-singlet conformal primaries that appear in the OPE $\phi_i \times \phi_j$ of two $O(N)$ fundamental operators, and impose a bound (that we hold fixed) on the second lowest dimension $\Delta_{\sigma^2}$ of such an operator.  Then varying the lowest scaling dimension $\Delta_\sigma$ of such an $O(N)$ singlet, we find an allowed region that exhibits the kink around $\Delta_{\sigma}\approx 2$, where the interacting theories of~\cite{Fei:2014yja,Fei:2014xta} were expected to exist.  For large values of $N$, we match the location of this kink with results coming from the large $N$ expansion.

There are several questions that we leave open and that we hope to come back to in the future.  Perhaps the most interesting one is how to determine precisely whether in $d=5$ there exists a critical value of $N$ below which the interacting $O(N)$ theory of~\cite{Fei:2014yja,Fei:2014xta} ceases to exist.  We noticed in Figure~\ref{fig:5dLowN} that under a reasonable assumption on $\Delta_{\sigma^2}$, the kink in the $(\Delta_\phi, \Delta_\sigma)$ plane disappears somewhere between $N=15$ and $N=22$.  It is tempting to conjecture that the critical value of $N$ lies in this range.  However, we noticed that in $d=5.95$, the kink corresponding to the interacting $O(N)$ CFT persisted down to smaller values of $N$ than what was expected from the $\epsilon$-expansion, and the same could be true in $d=5$ as well.  We therefore make a more conservative conjecture that, in $d=5$, there is no interacting $O(N)$ CFT for $N<15$.\footnote{This value is close to the estimate $N_\text{crit} \approx 14$ obtained in \cite{Fei:2014xta} from extrapolating the $4+\epsilon$ expansion to $\epsilon = 1$.}  It would be very interesting to find a systematic way of determining the precise value of $N_\text{crit}$ in this case.  Perhaps one way to proceed would be to examine whether there are any qualitative changes in the spectrum of operators for a potential CFT that lives at the kink as one varies $N$ using, for instance, the method of \cite{ElShowk:2012hu, El-Showk:2014dwa}.

Along these lines, it is worth pointing out a similarity between our study and that of \cite{Golden:2014oqa}, where $\Z_2$-invariant CFTs were examined in $d<3$.  For $d \geq 3$, the upper bounds on the dimension $\Delta_\sigma$ of the lowest scalar that appears in the $\phi \times \phi$ OPE (where $\phi$ is the lowest dimension $\Z_2$-odd operator) exhibit a kink in the $(\Delta_\phi, \Delta_\sigma)$ plane corresponding to the Ising model.  Below $d =3$, this kink splits into two distinct kinks, and a careful examination of the operator spectrum near these kinks shows that, if these kinks were to correspond to CFTs, they would be of different nature than the Ising model in $d \geq 3$.  We did observe a second kink $d = 5.95$ in Figures~\ref{fig:N595global} and~\ref{fig:595Compare}, but it is currently unclear whether this second kink gets washed out as we increase the parameter $\Lambda$ in our numerics.  It would therefore be very interesting to increase $\Lambda$ further, perhaps using a different semi-definite programing solver from {\tt sdpa\_gmp}.  If the second kink does not get washed out, it is conceivable that its appearance could be correlated with the lack of unitarity that occurs at the critical value of~$N$.

While in this work we focused on a two-dimensional section in the space $(\Delta_\phi, \Delta_\sigma)$ for a fixed lower bound on $\Delta_{\sigma^2}$, it would be interesting to let this lower bound on $\Delta_{\sigma^2}$ vary and obtain a three-dimensional plot.  Alternatively, one can assume that in the $O(N)$-singlet sector there are only two relevant conformal primary operators, $\sigma$ and $\sigma^2$, and vary their dimensions to obtain a three-dimensional plot.  Preliminary exploration shows that a fixed $\Delta_{\sigma^2}$ section of such a plot coincides with the plots shown in this paper where we simply impose a lower bound on $\Delta_{\sigma^2}$.

Lastly, it would be desirable to extend the bootstrap studies we performed beyond using the crossing symmetry of a single four-point function.  In the $d=3$ Ising case, such an extension provided a much more constrained allowed region in the $(\Delta_\phi, \Delta_\sigma)$ plane that includes a small island around the Ising point.  We leave such a study in $d=5$ for future work.


\subsection*{Acknowledgments}
\label{s:acks}

We are particularly grateful to Simone Giombi  and Igor Klebanov for many insightful conversations. We are also grateful to Vinod Gupta and Sumit Saluja for their help with using the Princeton Physics Department Feynman Computing Cluster. SSP and RY also thank Luca Iliesiu, Filip Kos, David Poland, and David Simmons-Duffin for collaboration on a related project.  This work was supported in part by the US NSF under Grant No.~PHY-1418069.


\appendix
\section{Fixed Point Solutions in $d=6-\epsilon$ for $\epsilon\ll1$}
\label{epsilon}

We now give the approximate dimensions of $\phi$, $\sigma$, and $\sigma^2$ in the $6-\epsilon$ expansion for the values of $N$ we use in Figure~\ref{fig:N595global}.\footnote{We thank S.~Giombi and I.~Klebanov for help in generating these expansions.}  Recall that in $6-\epsilon$ dimensions, the Lagrangian  \eqref{lagrangian} has three fixed points that were refer to as the critical theory, Theory 2, and Theory 3.  They correspond, respectively, to the red dot, the black dot close to the red dot, and the other black dot in Figure~2 of~\cite{Fei:2014yja}.  The critical theory has only two relevant scalar operators, while the other two have three relevant scalar operators. 

Note that the critical theory and Theory 2 values have imaginary components that are nonnegligible for $N<N_{\text{crit}}\approx 1000$, while the Theory 3 values are real for all $N>0$. The scaling dimension $\Delta_{\sigma^2}$ has only been computed to order $O(\epsilon)$, and to this order it differs by a non-negligible amount from the large $N$ value. Since the large $N$ value of $\Delta_{\sigma^2}$ is known to order $O(1/N^3)$, and since the value for $d=5.95$ \eqref{sigma2largeN} seems converged for the large values of $N$ considered in this paper, thus we use the large $N$ value of $\Delta_{\sigma^2}$ instead of the $\epsilon\ll1$ expansion value.  

\begin{align}
N=600:\quad& \qquad \notag\\
\text{Critical}:& \qquad \notag\\
\Delta_{\phi}&= 2-(0.498178-0.000018 i) \epsilon -(0.00208299-0.00003561 i) \
\epsilon ^2 \notag\\
&\quad+(0.000225164-0.000106536 i) \epsilon ^3 \label{600critphi}\\
\label{600critsig}
\Delta_\sigma&= 2+(0.0837282-0.0301539
i) \epsilon -(0.0657586-0.0431323 i) \epsilon ^2 \notag\\
&\quad+(0.0000980+0.0342568 \
i) \epsilon ^3 \\
\Delta_{\sigma^2}&= 4-(0.202864-0.336716
i) \epsilon
\label{600critsig2}\\
\text{Theory 2}:& \qquad \notag\\
\Delta_{\phi}&=2-(0.498178+0.000018 i) \epsilon -(0.00208299+0.00003561 i) \
\epsilon ^2 \notag\\
&\quad+(0.000225164+0.000106536 i) \epsilon ^3 \label{600t3phi}\\
\label{600t3sig}
\Delta_\sigma&=2+(0.0837282+0.0301539
i) \epsilon -(0.0657586+0.0431323 i) \epsilon ^2\notag\\
&\quad+(0.0000980-0.0342568 \
i) \epsilon ^3 \\
\Delta_{\sigma^2}&=4-(0.202864+0.336716
i) \epsilon\label{600t3sig2} \\
\text{Theory 3}:& \qquad \notag\\
 \Delta_{\phi}&=2-0.498897 \epsilon -0.000547562 \epsilon ^2+0.000444379 \
\epsilon ^3 \label{600t2phi}\\
\label{600t2sig}
\Delta_\sigma&=2-0.0798213 \epsilon +0.0419642 \epsilon ^2+0.0954303 \epsilon
^3 \\
\Delta_{\sigma^2}&=4-1.31623 \epsilon
\label{600t2sig2} 
\end{align}

\begin{align}
N=1000:\quad& \qquad \notag\\
\text{Critical}:& \qquad \notag\\
 \Delta_{\phi}&=2-0.498960 \epsilon \
-(0.001094448-0.000041800
i) \epsilon ^2\notag\\
&\quad+(0.000050697+0.000189705 i) \epsilon ^3 \label{1000critphi}\\
\label{1000critsig}
\Delta_\sigma&=2+(0.0578740+0.0053953 i) \epsilon -(0.0432267+0.0465933 i) \
\epsilon ^2\notag\\
&\quad-(0.014584+0.174052
i) \epsilon ^3 \\
\Delta_{\sigma^2}&=4-(0.287985+0.079340 i) \epsilon\label{1000critsig2} \\
\text{Theory 2}:& \qquad \notag\\
\Delta_{\phi}&=2-0.498960 \epsilon \
-(0.001094448+0.000041800
i) \epsilon ^2\notag\\
&\quad+(0.000050697-0.000189705 i) \epsilon ^3 \label{1000t3phi}\\
\label{1000t3sig}
\Delta_\sigma&=2+(0.0578740-0.0053953 i) \epsilon -(0.0432267-0.0465933 i) \
\epsilon ^2\notag\\
&\quad-(0.014584-0.174052
i) \epsilon ^3 \\
\Delta_{\sigma^2}&=4-(0.287985-0.079340 i) \epsilon \label{1000t3sig2} \\
\text{Theory 3}:& \qquad \notag\\
\Delta_{\phi}&=2-0.499306 \epsilon \
-0.000383821 \epsilon ^2+0.000232945 \epsilon ^3 \label{1000t2phi}\\
\label{1000t2sig}
\Delta_\sigma&=2-0.0639239 \epsilon +0.0345723 \epsilon ^2+0.0748065
\epsilon ^3 \\
\Delta_{\sigma^2}&=4-1.26281 \epsilon
\label{1000t2sig2}
\end{align}

\begin{align}
N=1400:\quad& \qquad \notag\\
\text{Critical}:& \qquad \notag\\
 \Delta_{\phi}&=2-0.499263 \
\epsilon -0.000730931
\epsilon ^2-.000006959\epsilon
^3 \label{1400critphi}\\
\label{1400critsig}
\Delta_\sigma&=2+0.0338577 \
\epsilon -0.0357827 \epsilon
^2+0.000671719\
\epsilon ^3 \\
\Delta_{\sigma^2}&=4-0.117458
\epsilon
\label{1400critsig2} \\
\text{Theory 2}:& \qquad \notag\\
\Delta_{\phi}&=2-0.499287 \
\epsilon -0.000734725\epsilon ^2+0.0000427058 \epsilon ^3 \label{1400t3phi}\\
\label{1400t3sig}
\Delta_\sigma&=2+0.0579004 \epsilon ^2-0.0348804\epsilon ^3 \\
\Delta_{\sigma^2}&=4-0.537913\epsilon\label{1400t3sig2} \\
\text{Theory 3}:& \qquad \notag\\
\Delta_{\phi}&=2-0.499491\
\epsilon -0.000297667 \epsilon ^2+0.000152964\epsilon ^3 \label{1400t2phi}\\
\label{1400t2sig}
\Delta_\sigma&=2.00000-0.0549757 \epsilon +0.0302051 \epsilon ^2+0.0634041
\epsilon ^3 \\
\Delta_{\sigma^2}&=4-1.23113 \epsilon
\label{1400t2sig2}
\end{align}

\bibliographystyle{ssg}
\bibliography{5dON}

\end{document}